\documentclass{article}
\usepackage{spconf,amsmath,graphicx,enumitem}
\usepackage{epsfig,amssymb,amsmath}
\usepackage{xfrac}

\newcommand{\R}{\mathbb{R}}

\title{A Vocoder-free WaveNet Voice Conversion with Non-Parallel Data}
\name{Xiaohai Tian$^{1,2}$, Eng Siong Chng$^{2}$ and Haizhou Li$^1$\thanks{This paper is submitted to INTERSPEECH 2019.}}
\address{$^1$Department of Electrical and Computer Engineering, National University of Singapore \\
$^2$School of Computer Engineering, Nanyang Technological University, Singapore \\
{\small \tt eletia@nus.edu.sg, aseschng@ntu.edu.sg, haizhou.li@nus.edu.sg}
}
\begin{document}
\maketitle
\begin{abstract}


In a typical voice conversion system, vocoder is commonly used for speech-to-features analysis and features-to-speech synthesis. 
However, vocoder can be a source of speech quality degradation.
This paper presents a vocoder-free voice conversion approach using WaveNet for non-parallel training data. Instead of dealing with the intermediate features, the proposed approach utilizes the WaveNet to map the Phonetic PosteriorGrams (PPGs) to the waveform samples directly. In this way, we avoid the estimation errors caused by vocoder and feature conversion.
Additionally, as PPG is assumed to be speaker independent, the proposed method also reduces the feature mismatch problem in WaveNet vocoder based approaches.
Experimental results conducted on the CMU-ARCTIC database show that the proposed approach significantly outperforms the baseline approaches in terms of speech quality.

\end{abstract}

\section{Introduction}

Voice conversion (VC) aims to modify the source speaker's voice to sound like that of the target speaker without changing the content information. 
The challenge is to transform the speaker identity while maintaining the speech quality.
Various techniques have been proposed to convert spectral feature for speaker identity conversion.
Among them, Gaussian mixture model (GMM)~\cite{stylianou1998continuous, kain1998spectral,toda2007voice,benisty2011voice} is one of the most popular methods, where the spectral feature is transformed by a statistical parametric model. 
However, it is known the GMM method doesn't capture the spectral details thus suffers from over-smoothing problem~\cite{toda2007voice,erro2010voice}.
To address these problems, frequency warping~\cite{erro2010voice, godoy2012voice, tian2014correlation, tian2015sparse} and exemplar based methods~\cite{takashima2013exemplar, wu2014exemplar} are also studied.
More recently, with good regression performance, neural network methods are widely used in VC task, e.g. deep neural network (DNN)~\cite{desai2009voice,chen2014voice,xie2014sequence}, long short-term memory (LSTM)~\cite{sun2015voice} and generative adversarial networks (GAN)~\cite{hsu2017voice, saito2018statistical}.

Despite the research progress, the quality of converted speech varies at run-time.
One reason is that most of the existing techniques perform the speaker identity conversion and speech reconstruction on the intermediate features analyzed by parametric vocoders. 
Conventional parametric vocoders (STRAIGHT~\cite{kawahara1999restructuring} and WORLD~\cite{morise2016world}) are designed based on certain assumptions, e.g. source filter model, time invariant linear filter. 
Additionally, to simplify mathematical formulation of the parametric model, some information, e.g. the phase information, are usually discarded.
As a result, the artifacts are introduced in both speech-to-features analysis stage and features-to-speech synthesis stage.  
To address the features-to-speech synthesis issue
a WaveNet vocoder~\cite{van2016wavenet, tamamori2017speaker, hayashi2017investigation, adiga2018use} is proposed to directly estimate the time domain waveform samples conditioned on input features.
Its effectiveness has been demonstrated in several voice conversion studies~\cite{kobayashi2017statistical, wu2018nu, liu2018wavenet, sisman2018voice} to replace the traditional vocoders for high quality speech generation.
However, these approaches continue to suffer from the feature mismatch problem between the training and generation of WaveNet vocoder.
As a result, undesired noise-like signals are observed in the WaveNet generated speech as reported in~\cite{kobayashi2017statistical, wu2018nu, wu2018collapsed}.


In this paper, we introduce a vocoder-free voice conversion approach using WaveNet for non-parallel training data, where the traditional parametric vocoder is not required for either intermediate spectral feature extraction or speech reconstruction. 
Inspired by~\cite{sun2016phonetic, tian2018average}, the proposed method first encodeds a speech signal into speaker independent (SI) feature representations, e.g. Phonetic PosteriorGrams (PPG)~\cite{sun2016phonetic, tian2018average}. Then, the WaveNet is trained to predict the corresponding time-domain speech signals with the SI features as the local conditioning.
At run-time, the same SI features extracted by given speech are used to drive the WaveNet to generate the converted speech. 
Note that, the conversion model is trained between the SI features and the corresponding time-domain speech signals of the same speaker.
Hence, the parallel data is not required for the proposed method.




This paper makes two main contributions,
\begin{itemize}[leftmargin=*,topsep=0pt]
\itemsep0em
\item Without using the parametric vocoder, the proposed approach prevents the feature extraction and speech reconstruction errors arising from the parametric vocoder;
\item Bypassing the the intermediate vocoder features for conversion, the proposed approach further reduces the recently proposed VC techniques with WaveNet vocoder.
\end{itemize}

\section{Voice Conversion with WaveNet Vocoder}

In this section, we discuss the advantages and limitations of the WaveNet vocoder based voice conversion techniques.

\subsection{WaveNet Vocoder}
\label{ssec:wavenet vocoder}

WaveNet vocoder~\cite{tamamori2017speaker} is a conditional WaveNet~\cite{van2016wavenet}. It can reconstruct the time-domain audio signals conditioned on the acoustic features extracted from traditional vocoders, e.g. aperiodicity, f0 and spectral features. 
Given a waveform sequence $x = [x_0,x_1,...,x_{T}]$ and the additional local conditioning input $\textbf{h}$, WaveNet vocoder can model the conditional distribution $p (x | \textbf{h})$ as follows:
\begin{equation}
\label{eq:cond_WaveNet}
p (x | \textbf{h}) = \prod_{t=1}^T p(x_t | x_1, x_2, \cdots, x_{t-1}; \textbf{h}).
\end{equation}

In order to model the long-range temporal dependencies of audio samples, an architecture based on dilated causal convolutions and a gated activation unit is proposed.
Deep residual learning framework is also utilized to speed up the convergence and train a deep model (e.g. 30 layers).
For the $i$-th residual block, the gated activation function is expressed as:
\begin{equation}
\label{eq:cond_WaveNet_gated}
\textbf{z}_i = \text{tanh}(\textbf{W}_{f,i} \ast \textbf{x} + \textbf{V}_{f,i} \ast \textbf{h}) \circ \sigma(\textbf{W}_{g,i} \ast \textbf{x} + \textbf{V}_{g,i} \ast \textbf{h}),
\end{equation}
where $\ast$ and $\circ$ denote the convolution and element-wise product operator respectively. $\textbf{W}$ and $\textbf{V}$ are the trainable convolution filters, $f$ and $g$ denote the filter and gate, respectively.

\subsection{The Limitations}

\begin{figure}[!htb]
\centering
\includegraphics[width=8.5cm]{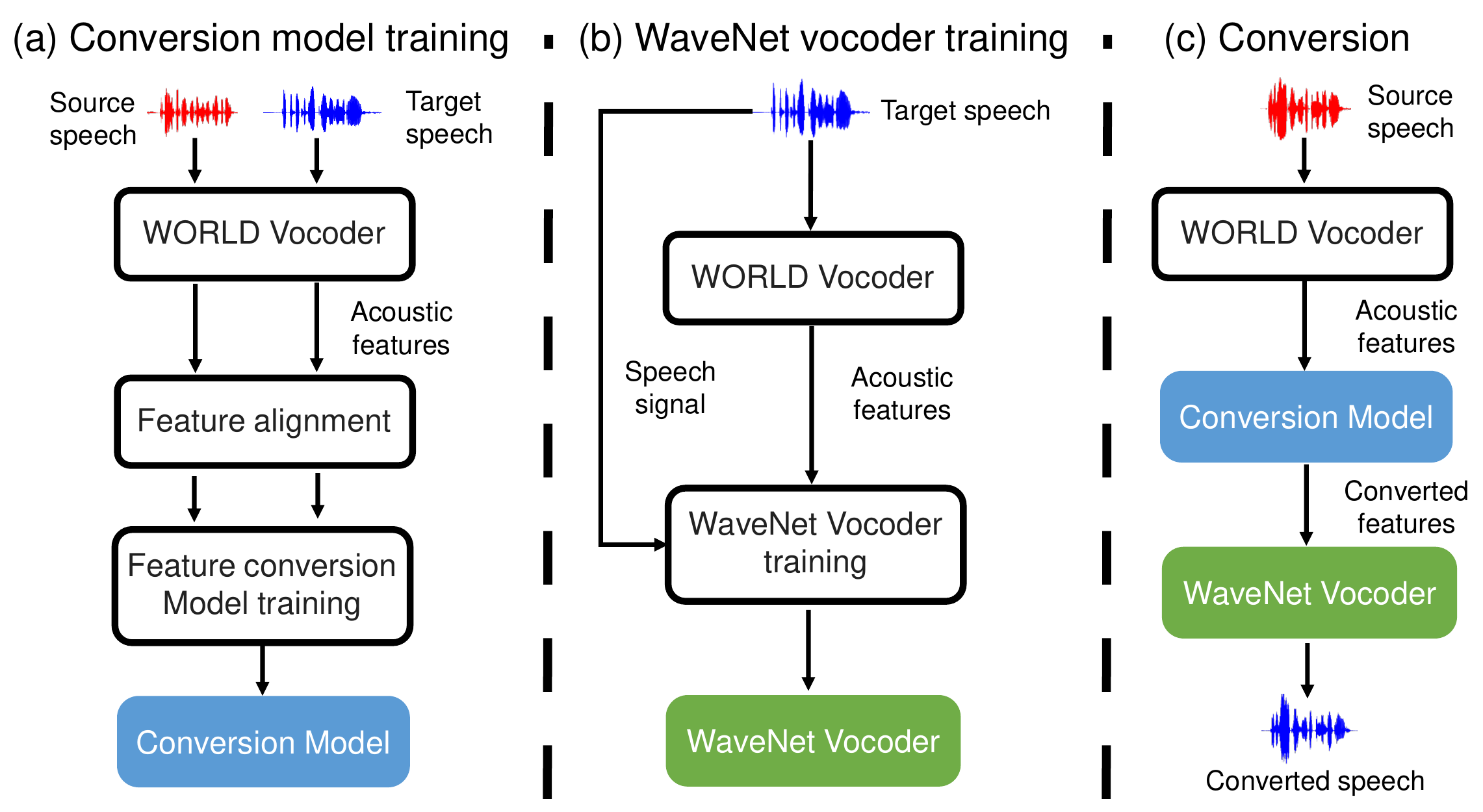}
\caption{Block diagram of WaveNet vocoder based voice conversion approach.}
\label{fig:WaveNet_vocoder}
\end{figure}

The WaveNet vocoder has been adopted in voice conversion tasks~\cite{kobayashi2017statistical, wu2018nu, liu2018wavenet, sisman2018voice} to replace the traditional vocoders for high quality speech generation. 
One of the successful example is proposed in~\cite{kobayashi2017statistical}, where the WaveNet vocoder is integrated in a GMM based VC framework.
Fig.~\ref{fig:WaveNet_vocoder} (a) and (b) show its conversion model and WaveNet vocoder training processes, while Fig.~\ref{fig:WaveNet_vocoder} (c) shows its conversion process.  
During training, two models are built. The GMM model is trained between the aligned source and target feature pairs for feature conversion. While, a WaveNet vocoder is trained with the acoustic features extracted from original target speech as the local conditioning input for speech generation.
At run-time, the acoustic features extracted by the traditional vocoders are first converted by GMM VC model. The converted features are then used as the additional input of the WaveNet vocoder to generate the converted speech.

While the WaveNet vocoder based VC is able to generate high quality speech, unstable problems of converted speech generation are reported in recent studies~\cite{kobayashi2017statistical, wu2018nu, wu2018collapsed}.
This is because the converted features used in run-time generation are very different to the original target features used for training, which results in the noise-like signals or irregular impulses in some speech segments~\cite{wu2018collapsed}.

\section{WaveNet Approach to Voice Conversion}

Phonetic PosteriorGrams (PPG) based voice conversion~\cite{sun2016phonetic, tian2018average} has been proposed to model the relationship between the PPG features to the corresponding acoustic features.
PPG is a sequence of probability vectors estimated with an automatic speech recognition (ASR) system.
As the ASR system is designed to generate the outputs invariant to the input speaker, the PPG feature is considered to be speaker independent.
Hence, it can be easily applied for voice conversion with non-parallel data.
In this paper, we investigate the effectiveness of using PPG as a local conditioning input of a WaveNet for vocoder free voice conversion. 
The proposed method does not rely on the intermediate features for speaker identity conversion.
Moreover, as PPG feature is considered to be speaker independent, the proposed system reduces the feature mismatch between WaveNet training and run-time stages. 

\begin{figure}[!htb]
\centering
\includegraphics[width=8cm]{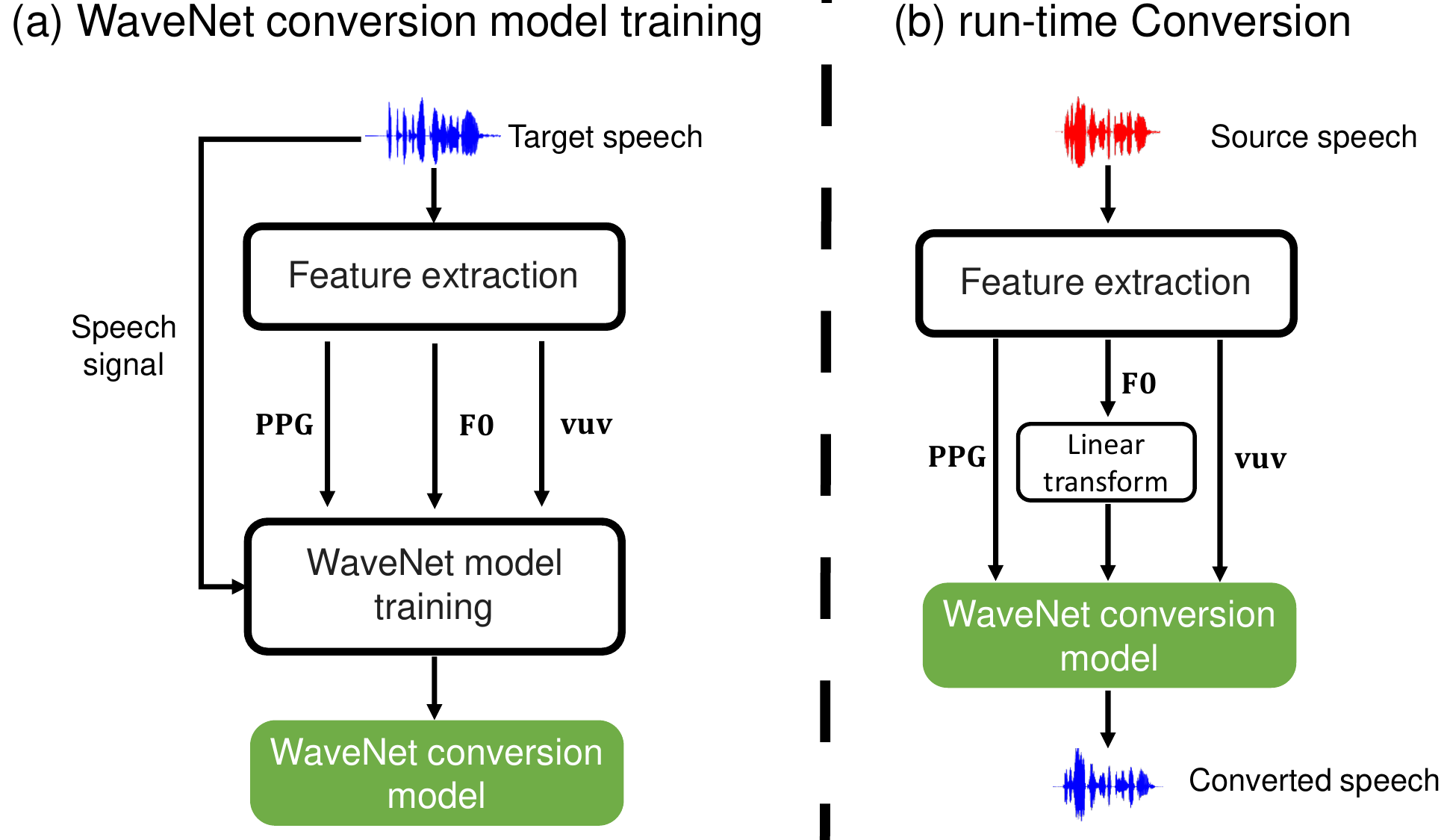}
\caption{Block diagram of the proposed vocoder-free voice conversion.}
\label{fig:PPG_WaveNet}
\end{figure}

The proposed framework is presented in Fig.~\ref{fig:PPG_WaveNet}, which consists of two steps: (a) WaveNet conversion model training and (b) run-time conversion. The details will be described as follows.

Fig.~\ref{fig:PPG_WaveNet}(a) shows the WaveNet conversion model training process. Given speech data of target speaker, we first extract PPGs $\mathbf{L} \in \R^{D \times N}$, where, $D$ and $N$ are the feature dimension and frame number respectively. 
In order to control the prosody of generated speech, $f0$ and voiced/unvoiced flag (vuv) features are also extracted, denoted as $\mathbf{F}0 \in \R^{1 \times N}$ and $\mathbf{F}_{\text{vuv}} \in \R^{1 \times N}$, respectively. 
To facility the WaveNet training, the PPGs, $f0$ and vuv are extended to match the temporal resolution of the time domain signals, denoted as $\widehat{\mathbf{L}} \in \R^{D \times T}$, $\widehat{\mathbf{F}0} \in \R^{1 \times T}$ and $\widehat{\mathbf{F}_{\text{vuv}}} \in \R^{1 \times T}$. 
Then, the local conditioning input $\textbf{h}$ in Eq.(\ref{eq:cond_WaveNet_gated}) can be expressed as $\mathbf{h} = [\widehat{\mathbf{L}}^\top, \widehat{\mathbf{F}0}^\top, \widehat{\mathbf{F}_{\text{vuv}}}^\top]^\top$.

At run-time (see Fig.~\ref{fig:PPG_WaveNet}(b)), given a source speech, we first extract the PPG, $f0$ and vuv features. 
A linear transformation is applied on the extracted $f0$, expressed as:
\begin{equation}
\widehat{{f0}_y} = \text{exp}( (\text{log} \widehat{{f0}_x} - {\mu}_{x}) {\frac{{\sigma}_y} {{\sigma}_{x}}} + {\mu}_{y}),
\label{eq:F0_mean_var_transformation}
\end{equation}
where ${\mu}_{x}$ and ${{\sigma}_{x}}$ are the mean and variance of the input source speech sample's $f0$ in logarithmic domain, respectively. ${\mu}_{y}$ and ${\sigma}_y$ are the mean and variance of the target speaker's $f0$ in logarithmic domain over all training samples. $\widehat{{f0}_y}$ is the converted $f0$ of the target speaker.
Then we adjust the temporal resolution of the PPG, converted $f0$ and vuv features and feed them into the trained WaveNet conversion model to generate converted speech.

\section{Experimental Setup}

\subsection{Database and feature extraction}
\label{ssec:data_feat}

The voice conversion experiments were conducted on the CMU-ARCTIC database~\cite{kominek2004cmu}.
Four speakers were selected consisting of two male speakers, \emph{bdl} and \emph{rms}, and two female speakers, \emph{slt} and \emph{clb}.
Intra-gender and inter-gender conversions were conducted between following pairs: 
\emph{rms} to \emph{bdl} (M2M), \emph{clb} to \emph{slt} (F2F), \emph{clb} to \emph{bdl} (F2M) and \emph{rms} to \emph{slt} (M2F).
500 utterances were used for training, another 20 non-overlap utterances of each speaker were used for evaluation. 

WORLD vocoder~\cite{morise2016world} was used to extract the 513-dimensional spectrum, 1-dimensional aperiodicity coefficients and $F_0$ with 5 ms frame step. Then 40-dimensional MCCs were calculated from the spectrum using Speech Signal Processing Toolkit (SPTK)~\footnote{https://sourceforge.net/projects/sp-tk/}.
The 42-dimensional phonetic posteriorgram (PPG) features were extracted by the PPG extractor trained on the Wall Street Journal corpus (WSJ)~\cite{paul1992design}. The detailed information can be found in~\cite{tian2018average}.
All the audio files were resampled at 16 kHz.

\subsection{Baselines and setup}
\label{baseline}

The details of reference systems and the proposed vocoder free voice conversion methods were introduced as follows.

\subsubsection{Reference Systems}
\begin{itemize}[leftmargin=*]
\itemsep0em
    \item \textbf{GMM-WORLD}: We implemented the joint-density Gaussian mixture model with maximum likelihood parameter conversion~\cite{kain1998spectral} for feature conversion. The WORLD vocoder was used for speech generation. The source and target MCC features were aligned using dynamic time warping (DTW)~\cite{sakoe1978dynamic}. Both static and its dynamic features were used in this implementation. The mixtures number of GMM is set to 128.
    \item \textbf{GMM(GV)-WORLD}: We use the same setting as GMM-WORLD, and the converted MCC features were enhaced by GV processing as proposed in \cite{toda2012implementation}.
	\item \textbf{GMM-WaveNet}: We use the same setting as GMM-WORLD with WaveNet vocoder for speech generation. 
	\item \textbf{GMM(GV)-WaveNet}: We use the same setting as GMM(GV)-WORLD with WaveNet vocoder for speech generation.
\end{itemize}

\subsubsection{The Proposed Vocoder-Free VC}
\begin{itemize}[leftmargin=*]
\itemsep0em
	\item \textbf{WaveNet-PPG}: The proposed WaveNet based voice conversion system with non-parallel data. The 42-dimensional PPG was used as the local condition of the WaveNet.
    \item \textbf{WaveNet-VC}: The proposed WaveNet based voice conversion system with non-parallel data. The 42-dimensional PPG, voiced/unvoiced flag and converted $f0$ were used as the local condition of the WaveNet. In total, the feature dimension was 44.
\end{itemize}

We trained the WaveNet vocoder and WaveNet conversion models for each target speaker.
Both WaveNet vocoder and WaveNet conversion models shared the same network architecture. The WaveNet consisted of 3 dilated residual blocks. 
Each residual block contained of 10 dilated causal convolution layers. In each block, the dilation started from 1 and exponentially increased by a factor of 2. 
The hidden units of residual connection and gating layers was set to 512, while the skip connection channels was set to 256.
The networks were trained using the Adam optimization method with a constant learning rate of 0.0001. The mini batch size was 15,000 samples and the training steps was set to 200,000.
The waveform sample values were encoded by 16 bits $\mu$-law.

\section{Evaluations}

\subsection{Objective evaluation}
\label{ssec:objective}

We conducted objective evaluation to assess the effectiveness of WaveNet-VC approach. The Root Mean Square Error (RMSE) was employed as the objective measure the distortion between target and converted speech. Magnitude features were extracted every 5ms with a window of 25ms.
For $j^{th}$ frame, the RMSE was calculated as:
$\text{RMSE[dB]} =
\sqrt{{\frac{1}{F} \sum_{f=1}^{F}} (20* \text{log}_{10} ( \frac{|Y(f)_{i}|} {|Y(f)_{i}^{\text{conv}}| }) )^2},$
where $Y(f)_{i}$ and $Y(f)_{i}^{\text{conv}}$ are the $i^{th}$ magnitude features of target and converted speech, respectively. $F$ is the total number of the frequency bins. A lower RMSE indicates the smaller distortion.


\begin{table} [h]
\caption{\label{table1} {\it Comparison of the Root Mean Square Errors (RMSEs) between the proposed vocoder-free VC and the reference systems.}}
\vspace{2mm}
\centerline{
\begin{tabular}{|c|c|c|c|c|c|}
\hline
Conversion Method &  Intra & Inter & Average \\
\hline
GMM-WORLD & 11.36 & 11.39 & 11.38\\
GMM(GV)-WORLD & 11.90 & 11.99 & 11.94\\
GMM-WaveNet & 13.52 & 13.46 & 13.49\\
GMM(GV)-WaveNet & 13.85 & 14.06 & 13.96\\
\hline
WaveNet-PPG & 14.32 & 14.89 & 14.61\\
\textbf{WaveNet-VC} & 13.62 & 13.84 & 13.73\\
\hline
\end{tabular}}
\end{table}

Table~\ref{table1} shows the RMSE results for all the baseline methods.
Firstly, we examine the effect of the $f0$ and voiced/unvoiced flag as a additional condition for WaveNet voice conversion.
It is observed that WaveNet-VC consistently outperforms the WaveNet-PPG in both intra- and inter-gender conversions.

Then, we further compare the performance of WaveNet-VC with other baseline methods.
We observe that the WaveNet-VC performs similar to the WaveNet vocoder baselines, with averaged RMSEs over all the testing pairs of 55.88 dB, 54.83 and 56.53 dB respectively.
The systems using WORLD vocoder outperform those with WaveNet vocoder. GMM-WORLD achieves the lowest RMSE of 46.89 dB. 

Note that, the objective metric evaluates the spectral distortion that reflects the how close the generated voice is to the target speech.
However, it is an indirect measurement. Typically, speech generated by traditional vocoders give a lower objective measure than that of WaveNet~\cite{tamamori2017speaker, adiga2018use}.
 
\subsection{Subjective evaluation}
  
AB preference tests and XAB tests were conducted to assess the speech quality and speaker similarity respectively.
In AB preference tests, each paired samples A and B were randomly selected from the proposed method and one of the baseline methods, respectively. Each listener was asked to choose the sample with better quality.
While, in XAB preference tests, X indicated the reference target sample, A and B were the converted samples randomly selected from the comparison methods. Noted that X, A and B have the same language content. The listeners were asked to listen to the samples, then decided  A and B which is closer to the reference sample or no preference.
For each test, 20 sample pairs were randomly selected from the 80 paired samples. 10 subjects participated in each tests.
Only the WaveNet vocoder based VC baselines, GMM-WaveNet and GMM(GV)-WaveNet were included in the listening tests.

%

\begin{figure}[!htb]
\centering
\includegraphics[width=8cm]{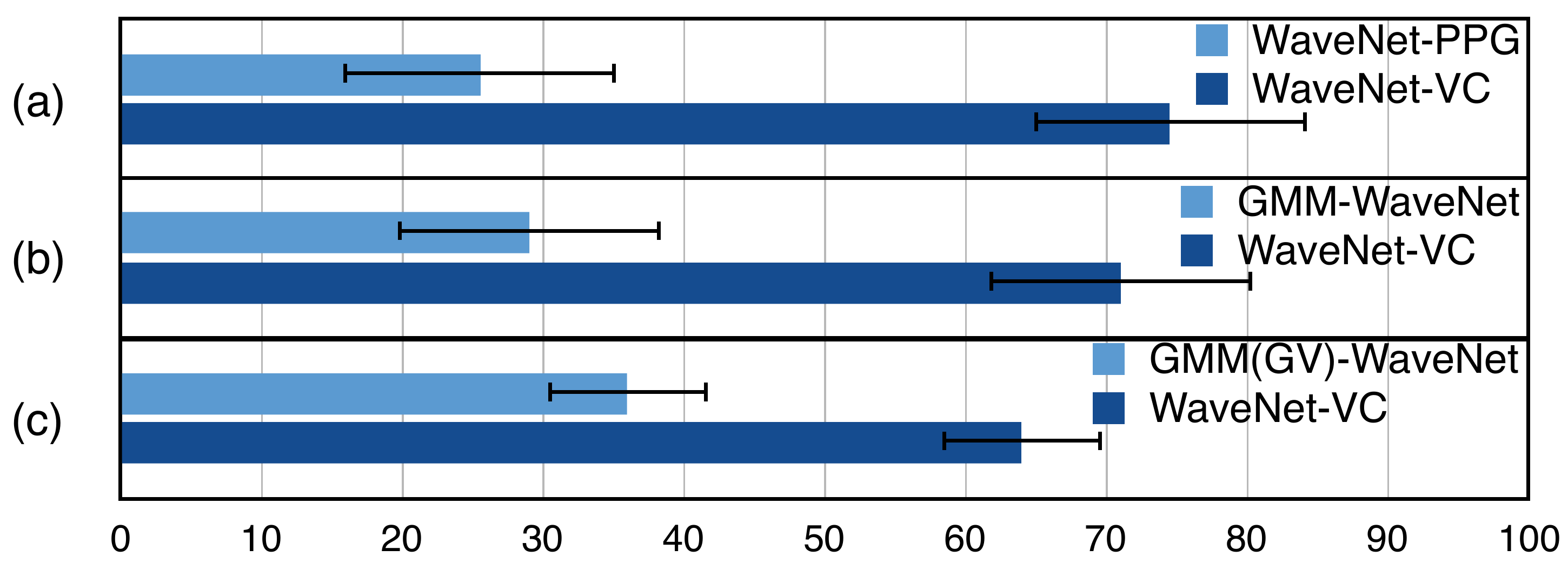}
\caption{Results of quality preference tests with 95\% confidence intervals for different methods.}
\label{fig:quality}
\end{figure}

The subjective results of quality preference tests are presented in Fig.~\ref{fig:quality}.
The results showed in Fig.~\ref{fig:quality} (a) suggests that the speech quality of WaveNet-VC significantly outperforms that of WaveNet-PPG.
Similar results are also observed in Fig.~\ref{fig:quality} (b) and Fig.~\ref{fig:quality} (b), which suggest that the proposed WaveNet-VC significantly outperforms GMM-WaveNet and GMM(GV)-WaveNet in terms of speech quality. 

\begin{figure}[!htb]
\centering
\includegraphics[width=8cm]{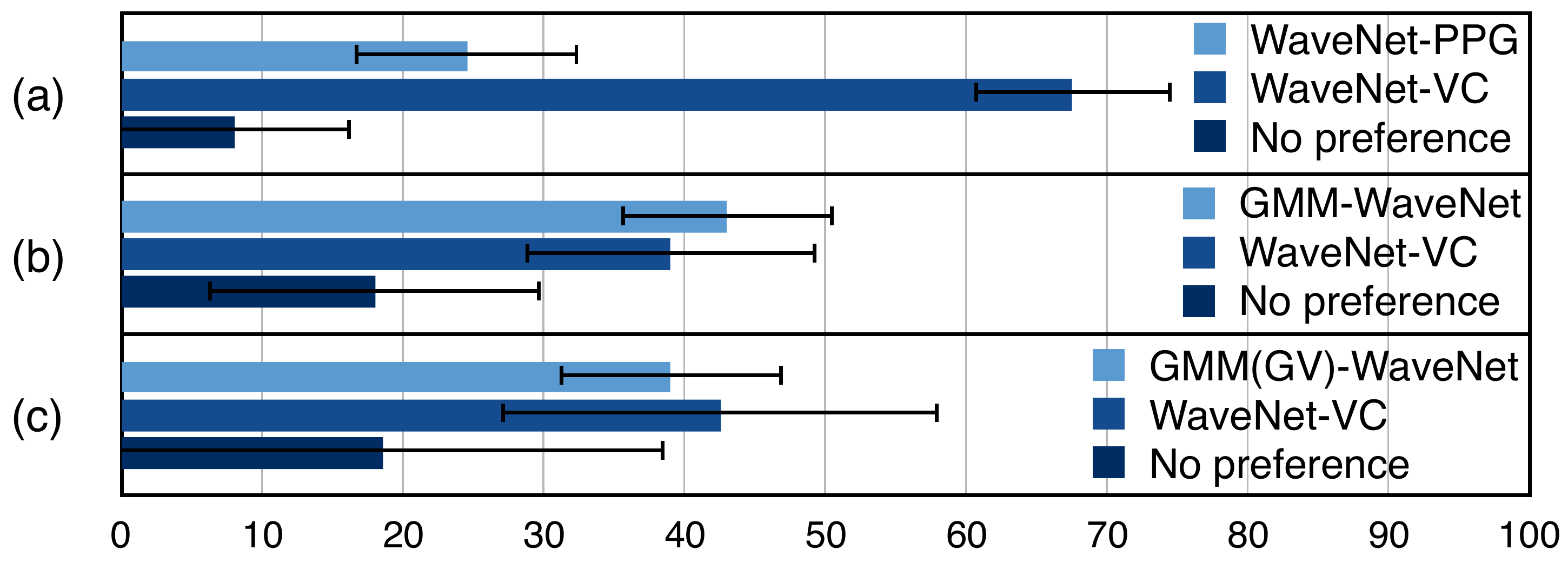}
\caption{Results of similarity preference tests with 95\% confidence intervals for different methods.}
\label{fig:similarity}
\end{figure}

The subjective results of speaker identity are presented in Fig.~\ref{fig:similarity}. We observe in Fig.~\ref{fig:similarity} (a) that WaveNet-VC consistently significantly outperforms WaveNet-PPG in terms of similarity.
While, in the experiments of WaveNet-VC vs. GMM-WaveNet and WaveNet-VC vs. GMM(GV)-WaveNet (see Fig.~\ref{fig:similarity} (b) and Fig.~\ref{fig:similarity} (c)), the identification rates fall into each other’s confidence intervals. This indicates that they are not significantly different in terms of speaker identity.

\section{Conclusions}
This paper presents a vocoder-free voice conversion approach using WaveNet for non-parallel data. 
The proposed approach does not rely on the vocoder features for conversion, which reduces the feature mismatch problem in WaveNet vocoder based approaches.
Experiment results show that the WaveNet-VC significantly outperforms the baseline methods in terms of quality, while maintain the speaker identity.


  \newpage
  \footnotesize
  \bibliographystyle{IEEEtran}
  
\bibliography{ICASSP2019}

\end{document}